# Post-buckling Solutions of Hyper-elastic Beam by Canonical Dual Finite Element Method

Kun Cai[1], David Y. Gao[2,3*] and Qing H. Qin[3]

1   College of Water Resources and Architectural Engineering, Northwest A&F University, Yangling, 712100, China
2   School of Science, Information Technology and Engineering, University of Ballarat, Mt Helen, Victoria 3353, Australia
3   Research School of Engineering, Australia National University, Canberra ACT 0200, Australia

* Corresponding author's email: d.gao@ballarat.edu.au (David Y. Gao).

**Abstract:** Post buckling problem of a large deformed beam is analyzed using canonical dual finite element method (CD-FEM). The feature of this method is to choose correctly the canonical dual stress so that the original non-convex potential energy functional is reformulated in a mixed complementary energy form with both displacement and stress fields, and a pure complementary energy is explicitly formulated in finite dimensional space. Based on the canonical duality theory and the associated triality theorem, a primal-dual algorithm is proposed, which can be used to find all possible solutions of this nonconvex post-buckling problem. Numerical results show that the global maximum of the pure-complementary energy leads to a stable buckled configuration of the beam. While the local extrema of the pure-complementary energy present unstable deformation states, especially. We discovered that the unstable buckled state is very sensitive to the number of total elements and the external loads. Theoretical results are verified through numerical examples and some interesting phenomena in post-bifurcation of this large deformed beam are observed.

## 1. Beam model and motivation

Large deformed cantilever beam to be studied is shown in Fig. 1, which is subjected to a distributed load $q(x)$ and axial compressive force $F$. To solve this problem using FEM, the beam is discretized with several beam elements, for the $e$-th element its two ends are marked as $A$ and $B$. The deflections of the two ends are $w_{Ae}$ and $w_{Be}$, respectively. The rotating angles of the two ends are $\theta_{Ae}$ and $\theta_{Be}$, respectively.

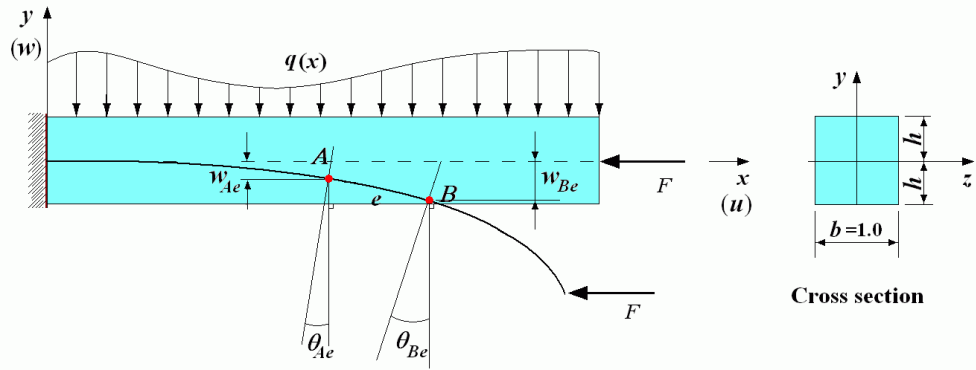

Fig. 1  Beam model (cantilever type)

In traditional nonlinear elastic beam theory, the stress in lateral direction (normal to the beam axis) is neglected and the governing equations can be considered as the one-dimensional von Karman model [1]:

$$\begin{aligned} \sigma_{x,x} &= 0 \\ EIw_{,xxxx} - \sigma_x w_{,xx} - q(x) &= 0 \end{aligned} \quad (1)$$

The first equation implies a constant stress field $\sigma_x$ which means that this von Karman beam model is actually a linear ordinary differential equation. However, if the thickness of a beam is quite large, e.g., thickness is of 10% to 20% of beam span, the deformation in lateral direction couldn't be neglected. To solve this problem mathematically, a nonlinear beam model was proposed by Gao in 1996 [2] which is controlled by the follow nonlinear differential equation

$$EI \cdot w_{,xxxx} - \alpha E w_{,x}^2 w_{,xx} + E\lambda w_{,xx} - f(x) = 0, \quad \forall x \in [0 \quad L] \quad (2)$$

where $\alpha = 3h(1-v^2) > 0$, $I = 2h^3/3$, $\lambda = (1+v)(1-v^2)F/E > 0$ and $f(x) = (1-v^2) \cdot q(x)$. The height of the beam is $2h$. $E$ is the elastic modulus of material and $v$ is the Poisson's ratio. $L$ is the length of the beam.

In this non-linear beam model, the following assumptions are considered: (a) cross-sections of beam are initially uniform along the beam axis and have a symmetry axis about which bending occurs; (b) cross sections remain perpendicular to the beam axis before and after deformation and shear deformation is ignored (Kirchhoff-Love hypothesis); and (c) the beam is under moderately large elastic deformation, i.e., $w(x) \sim h/L$ and $\theta \sim w_{,x}(x)$ ([2-4]). Comparing with the Euler–Bernoulli beam model, the present model ignores neither the stress nor the deformation of cross section in the lateral direction. The axial displacement field of the present beam model is given by the following equation [2]:

$$u_x = -\frac{1}{2}(1+\nu) w_{,x}^2 - \frac{\lambda}{2h(1+\nu)} \quad (3)$$

which implies that the axial deformation could be relatively larger. Therefore, this nonlinear beam model can be used for studying both pre and post buckling analysis for a large class of real-world problems in engineering and sciences ([5-9]).

The total potential energy $\Pi^p(w \in U_a \to R)$ of the beam associated with Eq.(2) is given by

$$\Pi^p(w) = \int_0^L \left( \frac{1}{2} EI \cdot w_{,xx}^2 + \frac{1}{12} E\alpha \cdot w_{,x}^4 - \frac{1}{2} E\lambda \cdot w_{,x}^2 \right) dx - \int_0^L f(x) \cdot w \, dx \quad (4)$$

where $U_a$ is the admissible deformation space of beam in which certain necessary boundary conditions are given. It is known from classic beam theory that the Euler buckling load can be defined as

$$F_{cr} = \inf_{w \in U_a} \frac{\int_0^L EI \cdot w_{,xx}^2 \, dx}{\int_0^L w_{,x}^2 \, dx} \quad (5)$$

If the end load $F < F_{cr}$, the beam is in an un-buckled state. The total potential energy $\Pi_p$ is convex and the non-linear differential equation (2) has only one solution. However, if $F > F_{cr}$, the beam is in a post-buckling state. In this case, the total potential energy $\Pi^p$ is non-convex and Eq. (2) may have at most three (strong) solutions [10] at each material point $x \in [0 \;\; L]$: two minimizers (one is a global minimizer and the other is a local minimizer), corresponding to the two stable buckling states, and one local maximizer, corresponding to an unstable buckling state. As we use numerical methods to solve the following non-convex variational equation

$$\delta \Pi^p(w, \delta w) = \delta \int_0^L \left( \frac{1}{2} EI \cdot w_{,xx}^2 + \frac{1}{12} E\alpha \cdot w_{,x}^4 - \frac{1}{2} E\lambda \cdot w_{,x}^2 \right) dx - \int_0^L f(x) \cdot w \, dx = 0, \quad (6)$$

we must encounter the non-uniqueness in a finite dimensional space. However, to find global optimal solution of a non-convex problem is usually NP-hard due to the lack of a global optimality condition [11]. It is well-known that for convex problems, the Hellinger–Reissner energy is a saddle-point functional, which connects each primal (potential energy) variational problem with an equivalent complementary dual problem. In contrast, for non-convex problems, the extremal property of the generalized Hellinger–Reissner principle and the existence of a purely stress-based complementary variational principle were two well-known debates existing in nonlinear elasticity for over 40 years ([12-16]). The first problem was partially solved by Gao and Strang (1989) [17]. By introducing a so-called complementary gap function, they recovered a broken symmetry in nonlinear governing equations of large deformation problems, and they proved that this gap function provides a global optimality condition. The second open problem was solved by Gao ([3, 18]) and a pure complementary energy variational principle was first proposed in both nonlinear beam theory [3] and general nonlinear elasticity [19]. A general review on this history was given in [20].

In the work by Gao ([2, 3]), a canonical dual transformation was presented, i.e.,

$$\sigma = \frac{E\alpha}{3} w_{,x}^2 - E\lambda \quad . \tag{7}$$

By using this canonical dual stress, the generalized total complementary energy of the beam $\Xi : U_a \times S_a \to R$ can be expressed as

$$\Xi(w,\sigma) = \int_0^L \left( \frac{1}{2} EI \cdot w_{,xx}^2 + \frac{1}{2} \sigma \cdot w_{,x}^2 - \frac{3}{4E\alpha} (\sigma + E\lambda)^2 - f(x) \cdot w \right) dx, \tag{8}$$

where $U_a, S_a$ are admissible spaces of deflection $w$ and the dual stress $\sigma$, respectively.

The Gao-Strang complementary gap function [17] for this beam model is defined as

$$G(w,\sigma) = \int_0^L \left( \frac{EI}{2} w_{,xx}^2 + \frac{\sigma}{2} w_{,x}^2 \right) dx. \tag{9}$$

Clearly, if the beam is subjected an extensive axial load $F$, the stress field $\sigma$ should be positive all over the domain and the gap function is convex in the displacement field $w$. In this case, the generalized complementary energy $\Xi(w,\sigma)$ is a saddle functional and the total potential energy $\Pi^p(w)$ is strictly convex, which leads to a unique global stable solution. If the axial load $F$ is a compressive force, the stress field $\sigma$ should be negative over $[0, L]$. However, as long as the axial stress is less than the Euler (pre-) buckling load, the gap function should be positive and the total potential energy is still convex. Therefore, this positive gap function provides a global stability criterion for general large deformation problems [17]. Seven years later it was discovered that the negative gap function can be used to identify the biggest local extrema in post-buckling analysis [3]. It turns out that a so-called triality theory was proposed by Gao in 1997. Furthermore, a pure complementary energy principle for finite elasticity theory was established in 1999 [18]. Since then, the canonical duality theory was gradually developed [21]. This theory is composed mainly of 1) canonical dual transformation, 2) a complementary-dual variational principle, and 3) the triality theory. Detailed information on this theory and its extensive applications in nonconvex mechanics and global optimization can be found in the monograph [21] as well as the review article [11].

Based on the Gao-Strang generalized complementary energy, the pure complementary energy of this nonlinear beam can be obtained by

$$\Pi^d(\boldsymbol{\sigma}) = \{\Xi(\boldsymbol{w},\boldsymbol{\sigma}) | \delta_w \Xi(\boldsymbol{w},\boldsymbol{\sigma}) = 0\} \quad (10)$$

which is defined on a statically admissible space $S_a$. Then we have the following result.

**Theorem 1: Complementary-dual principle**

The complementary energy $\Pi^d(\boldsymbol{\sigma})$ is canonical dual to the total potential energy $\Pi^p(\boldsymbol{w})$ in the sense that if $(\overline{\boldsymbol{w}}, \overline{\boldsymbol{\sigma}}) \in U_a \times S_a$ is a critical point of $\Xi(\boldsymbol{w},\boldsymbol{\sigma})$, then $\overline{\boldsymbol{w}} \in U_a$ is a critical point of $\Pi^p(\boldsymbol{w})$, $\overline{\boldsymbol{\sigma}} \in S$ is a critical point of $\Pi^d(\boldsymbol{\sigma})$ and $\Pi^p(\overline{\boldsymbol{w}}) = \Xi(\overline{\boldsymbol{w}}, \overline{\boldsymbol{\sigma}}) = \Pi^d(\overline{\boldsymbol{\sigma}})$.

In computational mechanics, it is well-known known that the traditional FEMs are based on the potential variational principle, which produces upper bound approaches to the related boundary-valued problems. The dual finite element method is based on the complementary energy principle, which was original studied by Pian *et al* (1964, 1978, 2006) ([22-24]) and Belytschko *et al* (1968, 1970, 1975) ([25-27]) mainly for infinitesimal deformation problems (although the buckling analysis with finite prebuckling deformations had also been studied by Glaum, Belytschko and Masur in 1975) [27]. The mixed/hybrid FEMs are based on the generalized Hellinger–Reissner complementary energy principle, which have certain advantages for solving both elastic deformation problems ([29, 30]) and large-scale structural plastic limit analysis [28].

Mathematically speaking, numerical discretization for nonconvex variational problems should lead to global optimization problems which could possess many local extrema. It is well-known in computational science that traditional direct approaches for solving nonconvex minimization problems in global optimization are fundamentally difficult or even impossible. Therefore, most of nonconvex optimization problems are considered as NP-hard [11]. Unfortunately, this well-known fact in computer science and global optimization is not fully recognized in computational mechanics. It turns out that many local search finite element methods have been used for solving large deformation problems.

The purpose of this article is to bridge the existing gap between computational mechanics and global optimization by developing a canonical dual finite element method (CD-FEM) for large deformation of beam model in Eq.(2). This method has been successfully applied for solving nonconvex variational problems in phase transitions of solids governed by Landau-Ginzburg equation [31] and post-buckling analysis of the nonlinear beam [32]. Due to the piece-wisely constant stress interpolation and coarse mesh scheme, the CD-FEM proposed in [32] can be used mainly for finding post-buckled solution of a beam under simple lateral loads. In this paper, our purpose is to find all possible solutions in the post-buckling analysis of the nonlinear beam model with complex lateral loads. Based on the canonical duality theory developed in Gao (2000b), we will first show that by using independent cubic shape functions for deflection and linear interpolation for dual stress field, the pure complementary energy function can be explicitly formulated in finite dimensional space. By the triality theory proved recently [33], a canonical primal-dual algorithm is then proposed which can be used to find all possible post-buckling solutions. Both stable and unstable buckled states of the nonlinear beam are investigated using this CD-FEM and illustrated by several examples.

## 2. Pure Complementary Energy and Triality Theory

Suppose the domain $\Omega$ can be discretized into finite elements $\Omega_e$, and each element has two nodes. Each node has three unknown parameters, e.g., for node "$A$" of element "$e$" (see Fig. 1), they are deflection ($w_{Ae}$), rotating angular ($\theta_{Ae}$) and dual stress ($\sigma_{Ae}$), for node "$B$", they are $w_{Be}$, $\theta_{Be}$ and $\sigma_{Be}$. The deformation field and the dual stress field are approximated, separately,

$$w_e^h(x) = \boldsymbol{N}_w^T \cdot \boldsymbol{w}_e \tag{11}$$

$$\sigma_e^h(x) = \boldsymbol{N}_\sigma^T \cdot \boldsymbol{\sigma}_e \tag{12}$$

where $\boldsymbol{w}_e^T = \begin{bmatrix} w_{Ae} & \theta_{Ae} & w_{Be} & \theta_{Be} \end{bmatrix}$ is the nodal displacement vector of the $e$-th element, $\boldsymbol{\sigma}_e^T = \begin{bmatrix} \sigma_{Ae} & \sigma_{Be} \end{bmatrix}$ is the nodal dual stress element. Their shape functions are as following

$$\boldsymbol{N}_w = \begin{bmatrix} \dfrac{1}{4}(1-\xi)^2(2+\xi) \\ \dfrac{L_e}{8}(1-\xi)^2(1+\xi) \\ \dfrac{1}{4}(1+\xi)^2(2-\xi) \\ \dfrac{L_e}{8}(1+\xi)^2(\xi-1) \end{bmatrix} \tag{13}$$

$$\boldsymbol{N}_\sigma^T = \dfrac{1}{2}\begin{bmatrix}(1-\xi) & (1+\xi)\end{bmatrix} \tag{14}$$

The generalized total complementary energy given in Eq. (8) can be expressed in discretized

form as the following

$$\Xi^h\left(\{w_e, \sigma_e\}\right) = \sum_{e=1}^{m-1}\left(\frac{1}{2}w_e^T \cdot G_e(\sigma_e) \cdot w_e - \frac{1}{2}\sigma_e^T \cdot K_e \cdot \sigma_e - \lambda_e^T \cdot \sigma_e - f_e^T \cdot w_e - c_e\right)$$
$$= \frac{1}{2}w^T \cdot G(\sigma) \cdot w - \frac{1}{2}\sigma^T \cdot K \cdot \sigma - \lambda^T \cdot \sigma - f^T \cdot w - c \qquad (15)$$

where $m$ is the number of nodes in beam, $w \in R^{2m}$, $\sigma \in R^m$ are nodal deflection and dual stress vectors, respectively; $\lambda \in R^m$, $f \in R^{2m}$, and $c \in R$ are defined by

$$\lambda_e = \int_{\Omega_e}\left(\frac{3}{2\alpha}\lambda N_\sigma(x)\right)dx = \frac{3\lambda L_e}{4\alpha}\begin{bmatrix}1\\1\end{bmatrix} \qquad (16)$$

$$f_e = \int_{\Omega_e}(f(x)N_w)dx = \int_{-1}^{1}\left(f\left(x = \frac{L_e}{2}(\xi+1)\right)N_w\right)\frac{L_e}{2}d\xi \qquad (17)$$

$$c_e = \int_{\Omega_e}\left(\frac{3}{4\alpha}\lambda^2 E\right)dx = \frac{3}{4\alpha}\lambda^2 E L_e \qquad (18)$$

The compliance matrix $K \in R^m \times R^m$ is composed by element matrices

$$K_e = \int_{\Omega_e}\left(\frac{3}{2E\alpha}N_\sigma \cdot N_\sigma^T\right)dx = \frac{L_e}{4E\alpha}\begin{bmatrix}2 & 1\\1 & 2\end{bmatrix} \qquad (19)$$

and the Hessian matrix of the gap function $G(\sigma) \in R^{2m} \times R^{2m}$ is obtained by assembling the following matrix in each element

$$G_e(\sigma_e) = \int_{\Omega_e}\left(EI \cdot N_w'' \cdot (N_w'')^T + ((N_\sigma)^T \cdot \sigma_e) \cdot N_w' \cdot (N_w')^T\right)dx$$
$$= \begin{bmatrix} g_{11} & g_{12} & g_{13} & g_{14}\\ & g_{22} & g_{23} & g_{24}\\ & & g_{33} & g_{34}\\ sym & & & g_{44}\end{bmatrix} \qquad (20)$$

where

$$\begin{cases}g_{11} = \frac{3}{2}t_1 + \frac{3}{5}t_2; \quad g_{12} = \frac{3L_e}{4}t_1 + \frac{L_e}{20}(t_2+t_3); \quad g_{13} = -\frac{3}{2}t_1 - \frac{3}{5}t_2\\ g_{14} = \frac{3L_e}{4}t_1 + \frac{L_e}{20}(t_2-t_3); \quad g_{22} = \frac{L_e^2}{2}t_1 + \frac{L_e^2}{30}(2t_2-t_3); \quad g_{23} = -g_{12}\\ g_{24} = \frac{L_e^2}{4}t_1 - \frac{L_e^2}{60}t_2; \quad g_{33} = g_{11}; \quad g_{34} = -g_{14}; \quad g_{44} = \frac{L_e^2}{2}t_1 + \frac{L_e^2}{30}(2t_2+t_3)\end{cases} \qquad (21)$$

and

$$t_1 = \frac{8EI}{L_e^3}; \quad t_2 = \frac{\sigma_{Be} + \sigma_{Ae}}{2}; \quad t_3 = \frac{\sigma_{Be} - \sigma_{Ae}}{2} \tag{22}$$

By the criticality condition

$$\delta \Xi^h(w, \sigma) = \delta \left( \frac{1}{2} w^T \cdot G(\sigma) \cdot w - \frac{1}{2} \sigma^T \cdot K \cdot \sigma - \lambda^T \cdot \sigma - f^T \cdot w - c \right)$$
$$= (G(\sigma) \cdot w - f) \cdot \delta w + \left( \frac{1}{2} w^T \cdot G_{,\sigma}(\sigma) \cdot w - K \cdot \sigma - \lambda \right) \cdot \delta \sigma = 0 \tag{23}$$

we obtain the following two equations:

$$G(\sigma) \cdot w - f = 0 \tag{24}$$

$$\frac{1}{2} w^T \cdot G_{,\sigma}(\sigma) \cdot w - K \cdot \sigma - \lambda = 0, \tag{25}$$

where $G_{,\sigma}$ stands for gradient of $G$ with respect to $\sigma$. Eq. (24) is actually a discretized equilibrium equation of (2), while Eq. (25) is the inversed constitutive relation. Let $S_a^h$ be a discretized feasible stress space such that $G$ is invertible for any given $\sigma \in S_a^h$. Then on the discretized feasible deformation space $U_a^h$, the displacement vector $w$ can be obtained by solving Eq. (26):

$$w = G^{-1}(\sigma) \cdot f \tag{26}$$

Substituting this into the generalized complementary energy Eq. (15), the discretized pure complementary energy can be explicitly given by

$$\Pi^d(\sigma) = -\frac{1}{2} f^T \cdot G^{-1}(\sigma) \cdot f - \frac{1}{2} \sigma^T \cdot K \cdot \sigma - \lambda^T \cdot \sigma - c \tag{27}$$

In order to identify both global and local extrema, we need the following subspaces:

$$S_a^{h+} = \left\{ \sigma \in R^{n_\sigma} \mid G(\sigma) \succ 0 \right\}, \tag{28}$$

$$S_a^{h-} = \left\{ \sigma \in R^{n_\sigma} \mid G(\sigma) \prec 0 \right\}, \tag{29}$$

Where $n_\sigma$ is the dimension of discretized stress field, correspondingly $n_w$ is the dimension of discretized deformation field. $G(\sigma) \succ 0$ means that the matrix $G$ is positive definite, while $G(\sigma) \prec 0$ stands for negative definite.

**Theorem 2: Triality theory**

Suppose $(\overline{w}, \overline{\sigma})$ is a critical point of $\Xi(w, \sigma)$.

**(1)** The critical point $\overline{w} \in U_a^h$ is a global minimizer of $\Pi^p(w)$ if and only if the critical point

$\bar{\sigma} \in S_a^{h+}$ is a global maximizer of $\Pi^d(\sigma)$, i.e.,

$$\Pi^p(\bar{w}) = \min_{w \in U_a^h} \Pi^p(w) \Leftrightarrow \max_{\sigma \in S_a^{h+}} \Pi^d(\sigma) = \Pi^d(\bar{\sigma}) \tag{30}$$

**(2)** The critical point $\bar{w} \in U_a^h$ is a local maximizer of $\Pi^p(w)$ if and only if $\bar{\sigma} \in S_a^{h-}$ is a local maximizer of $\Pi^d(\sigma)$.

$$\Pi^p(\bar{w}) = \max_{w \in U_a^h} \Pi^p(w) \Leftrightarrow \max_{\sigma \in S_a^{h-}} \Pi^d(\sigma) = \Pi^d(\bar{\sigma}) \tag{31}$$

**(3)** If $\bar{\sigma} \in S_a^{h-}$ and $n_w = n_\sigma$, then the critical point $\bar{w} \in U_a^h$ is a local minimizer of $\Pi^p(w)$ if and only if $\bar{\sigma}$ is a local minimizer of $\Pi^d(\sigma)$. If $\bar{\sigma} \in S_a^{h-}$ but $n_w > n_\sigma$ (the case studied in this paper), the vector $w = G^{-1}(\bar{\sigma}) \cdot f$ is a saddle point of $\Pi^p(w)$ on $U_a^h$, which is a local minimum only on a subspace of $U_a^h$ such that the dimension of $U_s$, $n_{w\_sub}$, equals $n_\sigma$, i.e.

$$\Pi^p(\bar{w}) = \min_{w \in U_s} \Pi^p(w) \Leftrightarrow \min_{\sigma \in S_a^-} \Pi^d(\sigma) = \Pi^d(\bar{\sigma}) \tag{32}$$

This theorem plays an important role in non-convex mechanics and global optimization. It was shown by Gao and Ogden that in nonconvex variational/boundary value problem of Ericksen's elastic bar, the global extremum solutions are usually nonsmooth which can't be captured by using any traditional Newton-type methods [34]. In the following sections, we will show that by this triality theorem, both global and local extrema of the post-buckling beam can be obtained.

### 3. Canonical Primal-Dual Algorithm and Flowchart

Based on the triality theory, a canonical primal-dual algorithm for finding all the post-buckling configurations (both stable and unstable post-buckled states) can be proposed as the following.

Step 1: Initiate parameters: $k=0$, vectors ($\sigma^{(0)}$) and matrices;

Step 2: calculate $w^{(k)} = G^{-1}(\sigma^{(k)}) \cdot f$ (Eq.(26)) using FEM;

Step 3: Using tri-duality theorem to find stress fields ($\sigma^{(k+1)}$) corresponding three different configurations:

    (a) stable post-buckled configuration (Global Maximum)

$$\max_{\sigma}\left\{\Xi(\boldsymbol{w}^{(k)},\boldsymbol{\sigma})\right\}$$
$$\text{subject to } \frac{1}{2}\left(\boldsymbol{w}^{(k)}\right)^T \cdot \boldsymbol{G}(\boldsymbol{\sigma}) \cdot \boldsymbol{w}^{(k)} > 0 \tag{33}$$

(b) local unstable configuration (Local Maximum)

$$\max_{\sigma}\left\{\Xi(\boldsymbol{w}^{(k)},\boldsymbol{\sigma})\right\}$$
$$\text{subject to } \frac{1}{2}\left(\boldsymbol{w}^{(k)}\right)^T \cdot \boldsymbol{G}(\boldsymbol{\sigma}) \cdot \boldsymbol{w}^{(k)} < 0 \tag{34}$$

(c) unstable buckled configuration (Local Minimum)

$$\min_{\sigma}\left\{\Xi(\boldsymbol{w}^{(k)},\boldsymbol{\sigma})\right\}$$
$$\text{subject to } \frac{1}{2}\left(\boldsymbol{w}^{(k)}\right)^T \cdot \boldsymbol{G}(\boldsymbol{\sigma}) \cdot \boldsymbol{w}^{(k)} < 0 \tag{35}$$

Step 4:   Convergence check:

If $\left\|\boldsymbol{\sigma}^{(k+1)} - \boldsymbol{\sigma}^{(k)}\right\|/\left\|\boldsymbol{\sigma}^{(k)}\right\| \leq 1.0 \times 10^{-9}$

    go to step 5

Else

    $k=k+1$

    If $k < k^*$

      go to step 5

    Else

      go to 2

    End

  End

Step 5:   Stop for post processing

The maximum iteration $k^*$ equals 100 in this algorithm.

## 4. Numerical examples

In the following examples, the beams have the same span $l = 1.0$m, height $2h = 0.1$m. We assume that the elastic modulus $E = 1000$Pa and the Poisson's ratio $\nu = 0.3$.

### 4.1 Example 1-mesh dependence study

Our first example is a beam ($L=1.0$m) subjected to a concentrated lateral force with and $f(x = 0.5) = (1 - \nu^2) \cdot q(x) = 1.0N$ (see Fig. 2). On the right end, the beam is subjected to a compressive force $\boldsymbol{F}$ and $\lambda = (1+\nu)(1-\nu^2)F/E = 0.003m^2$. To investigate the mesh dependency of our results, the beam is discretized with 10, 20, 30, 40, 50 and 60 elements, respectively.

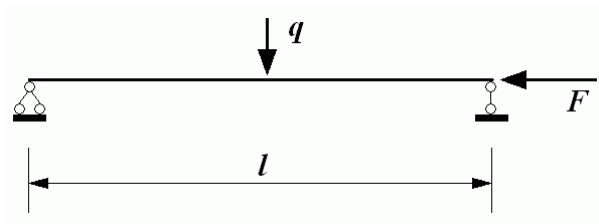

Fig. 2  Simply supported beam model

The curves with dark blue diamonds show the stable configurations (global maximum of $\Pi^d(\sigma)$) of the buckled beam with different number of elements. The maximum deflection at the centre of the beam is near 0.05m for different mesh schemes, which implies the mesh-independency of these stable configurations.

The curves with pink solid squares present the configurations (local maximum of $\Pi^d(\sigma)$) of the beam. Clearly, all deflections are almost close to zero. It means that the beam is nearly in a pure compressed deformation along the axial direction.

The curves with black triangles demonstrate the unstable configurations (local minimum of $\Pi^d(\sigma)$) of the beam. These curves are smooth but very with the total elements used. Therefore, it shows that these local unstable solutions are mesh-dependent. Especially, this unstable solution does not appear till the total number of elements (NE) > 20 (see Fig 3)**.** The reason for this mesh-dependency is that this local unstable solution is a saddle point of the nonconvex total potential energy which is sensitive to the discretization schemes of the original infinite dimensional problem. This result shows that the triality theory plays an important role in post-buckling analysis.

From the above analysis we can see that, as the total number of elements (NE) > 30, all the solutions (especially the third type) converge to the certain values. Therefore, we will use NE = 40 elements with the same beam length for all the following examples.

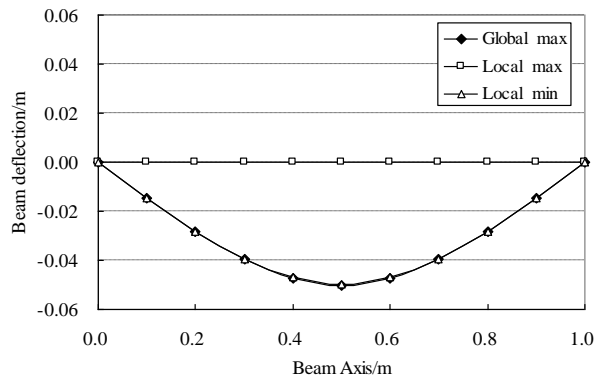
(a) NE=10

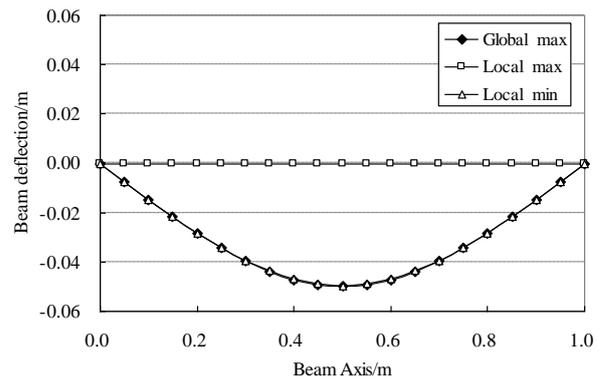
(b) NE=20

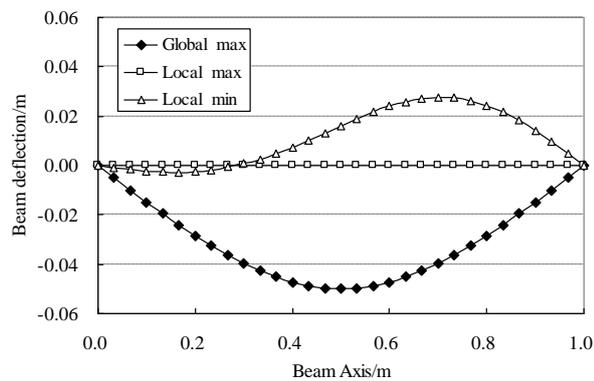
(c) NE=30

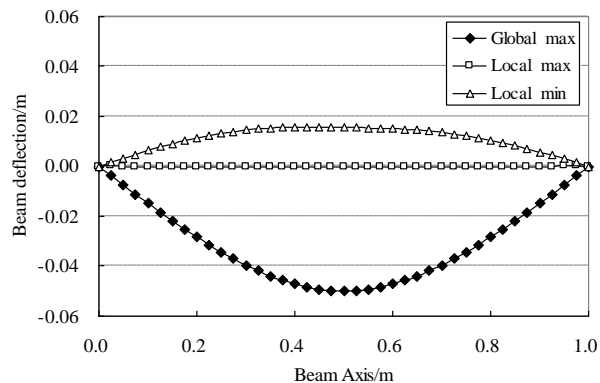
(d) NE=40

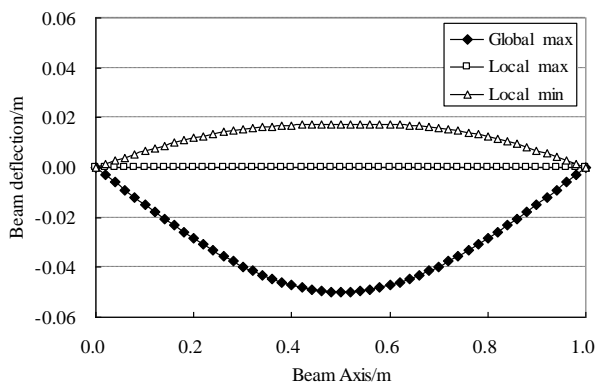
(e) NE=50

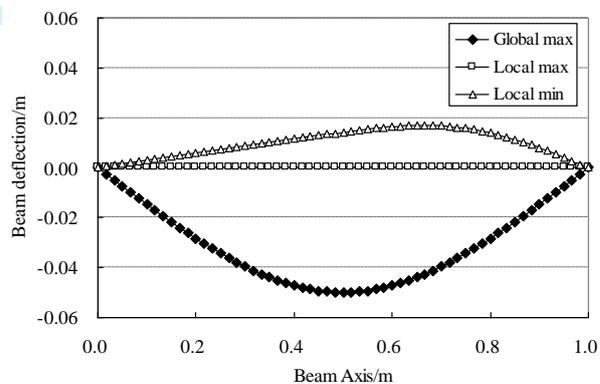
(f) NE=60

Fig. 3  Mesh-dependency of solutions of the buckled beam

### 4.2 Example 2-Complicated lateral load

Let us to consider a clamped/clamped beam as shown in Fig. 4. The lateral load is assumed to be piece wise uniform pressures ($q(x)$) such that $f(x) = (1-v^2) \cdot q(x) = 0.1 N/m$. The left end is fixed, while the deflection at the right end is specified as zero and there is a compressive force $F$. Two cases are considered for the end load: (a) $\lambda = (1+\nu)(1-\nu^2)F/E = 0.001m^2$; (b) $\lambda = 0.05m^2$.

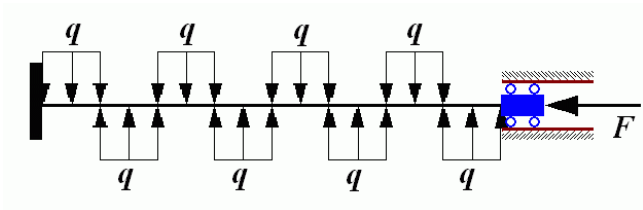

Fig. 4  Clamped/simply supported beam model

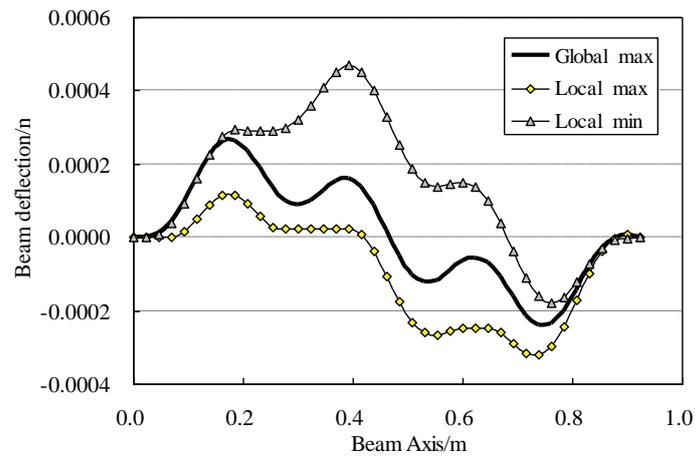

Fig. 5  Three post-buckling configurations of beam as $\lambda = 0.01m^2$

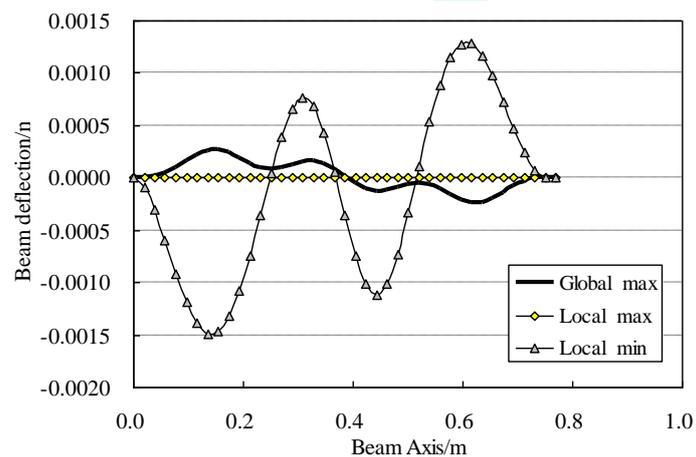

Fig. 6  Three post-buckling configurations of beam as $\lambda = 0.03m^2$

Since the lateral load is not uniformly distributed, this example can be used to study more interesting phenomena of the post-buckling configurations subjected to a combination of the lateral load ($q(x)$) and the end compression ($F$).

Fig. 5 and Fig. 6 present the three post-buckling configurations of the beam with $\lambda = 0.01 m^2$ and $\lambda = 0.01 m^2$, respectively. We can see that the stable buckled beam (global max of $\Pi^d(\sigma)$) is rotationally symmetric with respect to the beam center, but the other two unstable solutions are not symmetrical and are sensitive to the axial load, the bigger *F*, the larger axial deformation. Fig 6 shows the beam can have a deformation of over 20% of its length. This fact shows that the present method can be used for large deformed post-buckling analysis.

### 4.3 Example 3-Different lateral loads

Now let us consider a simply supported beam as shown in Fig. 7. The right end of the beam is subjected to a compressive force *F* such that $\lambda = (1+\nu)(1-\nu^2)F/E = 0.01 m^2$. Three types of lateral loads are considered: (a) a concentrated force $f(x=0.5l) = (1-\nu^2)\cdot q = 0.1N$ on the center of the beam, (b) a uniformly distributed load $f(x) = 0.1 N/m$, and (c) piece-wise uniform load $f(x) = 0.1 N/m$.

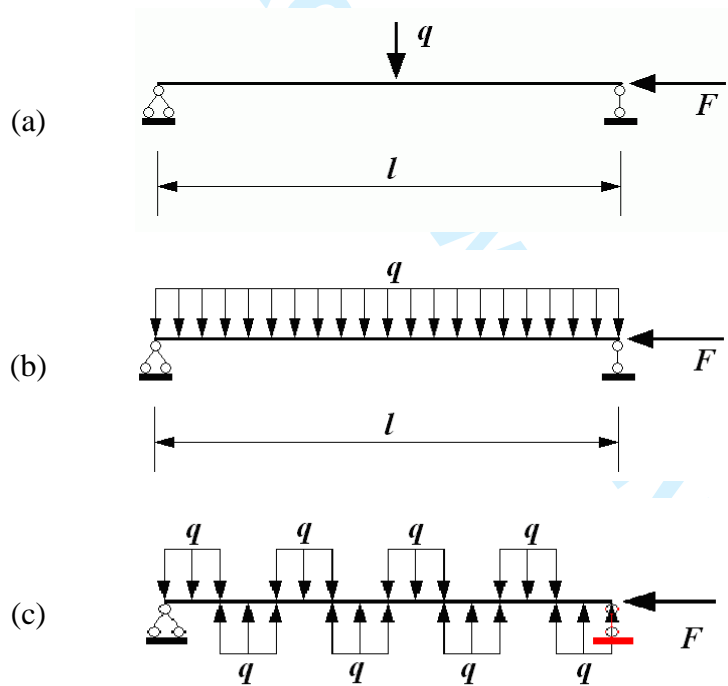

Fig. 7 Simply supported beam model

In this example, sensitivity of the post-buckling deformation with the lateral loads is investigated. Fig. 8 and Fig 9 present stable and unstable post-buckling deformations of the beam under concentrated and uniformed forces, respectively. Clearly, the stable buckled states are similar, but not the unstable buckled states. While for the periodic lateral load, the stable and unstable post-buckled configurations are similar in shapes but different in magnitudes (see Fig 10), which is reasonable since the stable buckled state is a global minimal solution to the nonconvex problem, but the unstable buckling state is only a stationary point of the total potential.

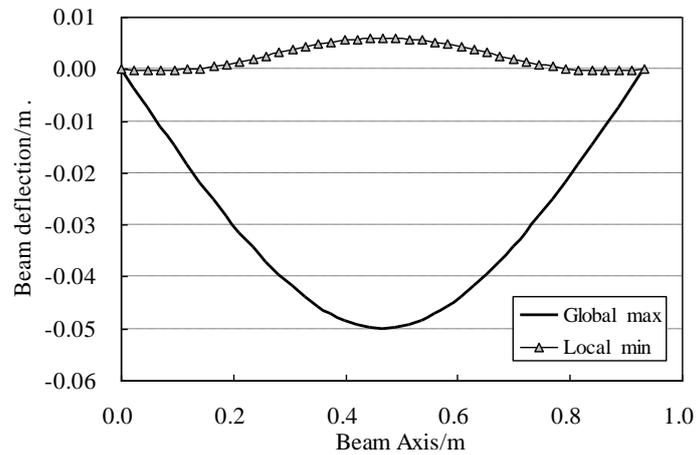

Fig. 8  Post-buckling configurations of the beam in case a)

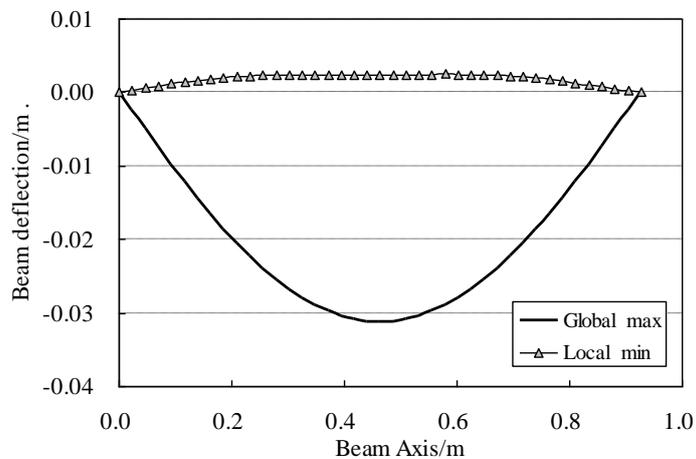

Fig. 9  Post-buckling configurations of the beam in case b)

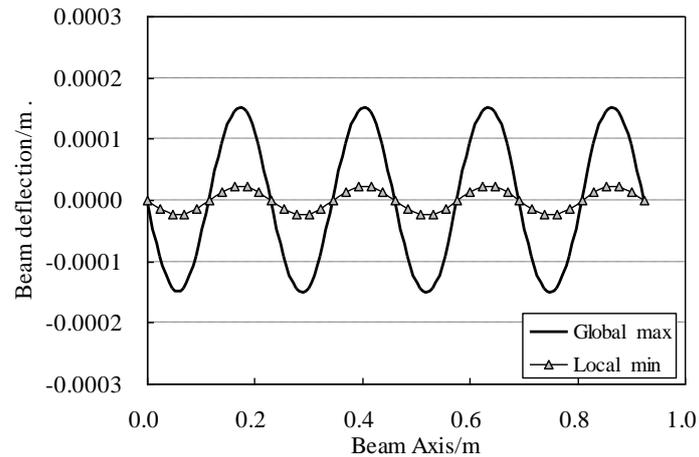

Fig. 10 Post-buckling configurations of the beam in case c)

**4.4 Example 4-Coupling of lateral load and axial compression**

Finally, we show some interesting phenomena of the unstable post-buckling state (i.e. the local minimum of $\Pi^d(\sigma)$) for a simply supported beam (see Fig 11) with different concentrated loads: $f(x=0.5l)=(1-v^2)\cdot q = 1.0N, 1.2N$, and $1.45N$, each one is combined with two different compressive forces: (a) $\lambda=(1+v)(1-v^2)F/E=0.003m^2$ and (b) $\lambda=0.008m^2$. We discovered that this unstable solution is very sensitive to the external loads as expected.

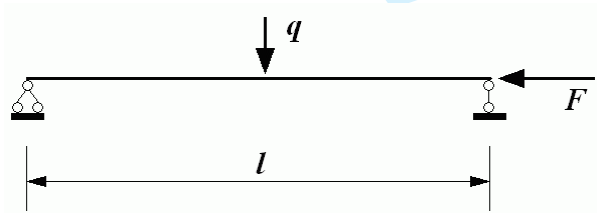

Fig. 11 Simply supported beam model

From Fig. 12 we can see that for the given compressive load $\lambda=0.003m^2$, the stronger is the concentrated load $f$, the smaller is the center deflection of the unstable buckled state. This case is easy to understand as the strong concentrated load should push down the unstable buckled beam. However, this phenomenon turns to the opposite way if the compressive load is increased to $\lambda=0.008m^2$ (see Fig 13). We believe that these phenomena need to have detailed study by both numerical simulations and experiments. Carefully observation shows that all these unstable post-buckled configurations are not symmetric to the center. For example, in Fig 12, we can see that the left peak (x, y)=(0.25m, 0.0372m) of the black curve (f=1.2*N*) is lower than the right peak (0.75m, 0.0466m). For the yellow curve (f=1.45 *N*), the left peak (0.2m, 0.048m) is higher than the right peak (0.8m, 0.0463m).   These non symmetric deformations are due to the non symmetrical boundary conditions.

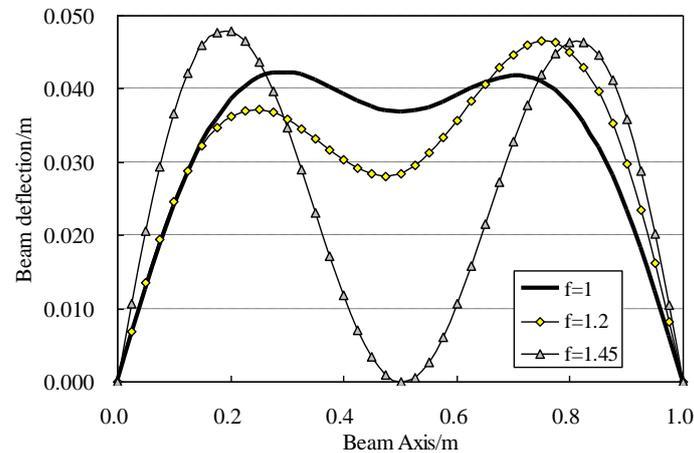

Fig. 12 . Unstable post-buckling deflections of the beam as $\lambda = 0.003 m^2$

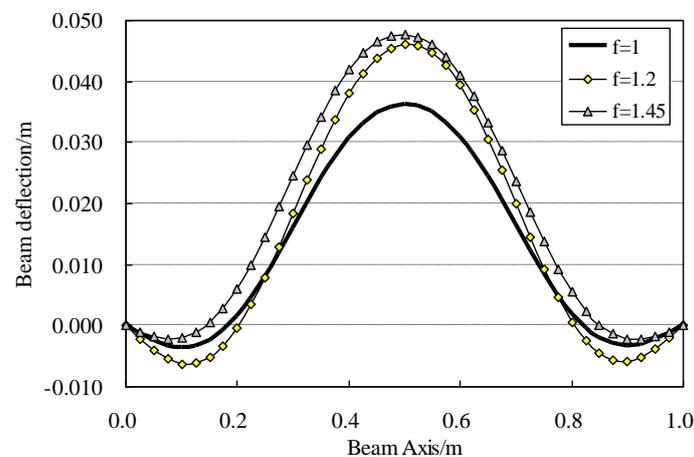

Fig. 13 Unstable post-buckling deflections of the beam with $\lambda = 0.008 m^2$

## 5. Conclusions

Based on the canonical duality theory, a mixed finite element method is proposed for solving a post-buckling problem of a large deformed nonlinear beam. By using independent shape functions for deflection and dual stress fields, a pure complementary energy function is explicitly formulated in finite dimensional space.   Combining this pure complementary energy with the triality theory, a canonical primal-dual algorithm is proposed which can be used to find not only the stable buckled solution, i.e. the global minimal of the nonconvex total potential, but also the unbuckled state and the unstable buckled state. Our numerical results show that the stable buckled state is indeed stable but the unstable post-buckled solution is very sensitive to both the total number of meshes and the external loads. Some interesting phenomena are discovered which deserve to have future study.

## Acknowledgements

Financial supports by US Air Force Office of Scientific Research (Grant No. FA9550-10-1-0487), the National Natural Science Foundation of China (Grant No. 50908190), Youth Talents Foundation of Shaanxi Province (Grant No. 2011kjxx02), and the Human Resource Foundation of Northwest A&F University (Grant No. QN2011125) are fully acknowledged.